\documentclass[AMS,STIX1COL]{WileyNJD-v2}

\articletype{Paper}

\begin{document}

\title{Sensitivity analysis and experimental evaluation of PID-like continuous
sliding mode control}

\author[1]{Michael Ruderman*}

\author[2]{Johann Reger}

\author[3]{Benjamin Calmbach}

\author[4]{Leonid Fridman}

\authormark{RUDERMAN et AL}

\address[1]{\orgname{University of Agder}, \orgaddress{\state{P.B. 422, Kristiansand, 4604}, \country{Norway}}}

\address[2]{\orgname{Technische Universit\"{a}t Ilmenau}, \orgaddress{\state{98684 Ilmenau}, \country{Germany}}}

\address[3]{\orgname{Technische Universit\"{a}t Ilmenau}, \orgaddress{\state{98684 Ilmenau}, \country{Germany}}}

\address[4]{\orgname{Universidad Nacional Autonoma de
Mexico}, \orgaddress{\state{04510 Mexico City}, \country{Mexico}}}

\corres{*Corresponding address. \email{michael.ruderman@uia.no}}

%\presentaddress{This is sample for present address text this is
%sample for present address text}

%%%
\abstract[Summary]{Continuous higher order sliding mode (CHOSM)
controllers represent an efficient tool for disturbance rejection.
For the systems with relative degree $r$, CHOSM approaches provide
theoretically exact compensation of the matched Lipschitz
perturbation, ensuring the finite-time convergence to the $(r +
1)$-th sliding-mode set, by using only information on the sliding
output and its derivatives up to the order $(r-1)$. In this paper,
we investigate the disturbance rejection properties of a PID-like
CHOSM controller, as the simplest and intuitively clear example
which incorporates nonlinear actions on the output error, its
derivative, and integration of its sign. We use the harmonic
balance approach and develop an analysis of propagation of the
matched Lipschitz perturbation through the control loop in
frequency domain. The resulted solution appears in form of the
Bode-like loci which depend also on the amplitude of harmonic
disturbances. Such amplitude-frequency characteristics allow
certain comparability with standard disturbance sensitivity
functions of a linear PID-controlled system in frequency domain.
Also a simple and straightforward design procedure for the robust
linear PID controller targeting the second-order system plants
under investigation is provided for benchmarking. Additional
(parasitic) actuator dynamics, which can lead to self-induced
steady oscillations, i.e. chattering, is ditto respected. A
detailed experimental case study, accomplished on an
electro-mechanical actuator in the laboratory setting, highlight
and make the pros and cons of both PID and CHOSM controllers well
comparable for a broadband disturbance rejection.}

\keywords{Sliding mode control, harmonic balance analysis,
disturbance rejection, sensitivity analysis, robust control
design, higher order sliding mode}

\maketitle

%\footnotetext{\textbf{Abbreviations:} ANA, anti-nuclear
%antibodies; APC, antigen-presenting cells; IRF, interferon
%regulatory factor}

%\newtheorem{thm}{Theorem}
%\newtheorem{lem}[thm]{Lemma}
%\newtheorem{clr}{Corollary}
%\newdefinition{remark}{Remark}
%\newdefinition{exmp}{Example}
%\newdefinition{prop}{Proposition}
%\newproof{pf}{Proof}

%%%%%%%%%%%%%%%%%%%%%%%%%%%%%%%%%%%%%%%%%%%%%%%%%%%%%%%%%%%%%%%%%%%%%%%%%%%%%%%%
\section{Introduction}
\label{sec:1}

Sliding-mode controllers (SMC) are especially efficient for
disturbance rejection\cite{shtessel2014}. The main disadvantage of
SMC is the presence of the so-called chattering effect due to the
discontinuity used in a control law, see e.g.
\cite{boiko2003},\cite{boiko2005},\cite{shtessel2014}. Continuous
higher order sliding mode (CHOSM) algorithms, see e.g.
\cite{levant1993sliding},\cite{zamora2013control},\cite{moreno2016},
\cite{kamal2016},\cite{laghrouche2017},\cite{cruz2019},\cite{mercado2020},
\cite{moreno2020},\cite{perez2021}, are a class of homogeneous
sliding mode controllers capable to compensate Lipschitz
uncertainties and perturbations theoretically exactly, and
generating a continuous control signal instead of a discontinuous
one. These algorithms consist of a static homogeneous finite-time
controller for the nominal model of the system and a discontinuous
integral action, aimed at estimating and compensating for the
uncertainties and perturbations. CHOSM controllers can be seen as
an extension of the (classical) super-twisting algorithm (STA)
\cite{levant1993sliding}, ensuring the finite-time convergence to
the $(r + 1)$-th sliding-mode set for systems with a relative
degree $r$. CHOSM controllers are using only information on the
sliding output and its derivatives up to the order $(r-1).$ The
homogeneity weight of the output variables in the CHOSM
controllers is $r+1$, cf.
\cite{moreno2016},\cite{cruz2019},\cite{mercado2020},\cite{moreno2020},
and it results, correspondingly, in a chattering amplitude of the
order $r+1$ \cite{levant2010}.

A PID-like continuous sliding mode controller introduced in
\cite{zamora2013control}, \cite{cruz2019} is the simplest and
intuitively clear CHOSM algorithm. For the systems with relative
degree $r = 2$, it ensures the finite-time convergence to the
third-order sliding-mode set using only information on the sliding
output and its derivative. Moreover, when the actuator is fast
enough, cf. \cite{perez2019}, \cite{perez2019b}, the chattering
effects caused by the discontinuity and discretization can be
strongly attenuated. To generate the synthesis rules for PID-like
controllers the authors of \cite{perez2021} used the harmonic
balance method and made the PID-like controller gains adjusting
the parameters of chattering. The latter is due to the presence of
parasitic (actuator) dynamics. Finally, an approach proposed in
\cite{perez2021} allows to minimize either the amplitude of
chattering or the energy needed to maintain the system with
relative degree two in a third-order sliding mode. The harmonic
balance-based analysis of PID-like controllers opened the door for
investigation of the propagation properties of CHOSM algorithms
that will be used in this paper. The corresponding contribution of
the paper is summarized as:

\begin{itemize}
\item an analysis of the PID-like CHOSM control based on its analytic
describing function, allowing to estimate the disturbance
sensitivity in frequency domain;

\item a comparison of the steady-state properties of the CHOSM
controller with a simple and straightforwardly designed robust
linear PID controller, that is based on the upper bound of the
disturbance sensitivity function;

\item a detailed experimental case study, accomplished on an
electro-mechanical actuator in the laboratory setting,
highlighting and making the pros and cons of both PID and PID-like
CHOSM controllers well illustrative.

\end{itemize}

The rest of the paper is organized as follows. In section
\ref{sec:2}, we address the PID-like CHOSM control
\cite{perez2021}. First we introduce the control problem. Then, we
summarize the CHOSM control with proportional, derivative and
integral terms in a closed-loop configuration with external
disturbances. Finally we perform the analysis based on harmonic
balance and propagation of harmonic disturbances through the
CHOSM-controlled system with additional parasitic actuator
dynamics. Section \ref{sec:3} is devoted to a robust design of a
linear PID control based on the sensitivity function. The
comparative experimental case study is reported in detail in
section \ref{sec:4}. Brief conclusions are drawn by the end in
section \ref{sec:5}.

%%%%%%%%%%%%%%%%%%%%%%%%%%%%%%%%%%%%%%%%%%%%%%%%%%%%%%%%%%%%%%%%%%%%%%%%%%%%%%%%
\section{PID-like Continuous Sliding Mode Control}
\label{sec:2}

In this section, the class of dynamic systems under consideration
is first stated together with the control problem. Then, the CHOSM
\cite{zamora2013control,cruz2019} targeted in this work is
summarized for the reader convenience. Finally, the analysis based
of harmonic balance equation and describing function of CHOSM
control is developed, which allows consideration in frequency
domain.

\subsection{Control problem}
\label{sec:2:sub:1}

Consider the class of perturbed second-order systems which can be
written, in time domain, as
\begin{equation}\label{eq:system}
\ddot{y}(t) = u(t) + d(t).
\end{equation}
The single measurable system state is the output $y(t)$, and the
available control channel is $u(t)$. The matched exogenous
disturbance $d(t)$ is assumed to be Lipschitz continuous, meaning
$|\mathrm{d}/\mathrm{dt} \, d(t)| \leq L$ while the upper bound $L
> 0$ is assumed to be known. Moreover, the control channel $u$ can
be subject to an additional first-order actuator dynamics with a
unity gain and a not-negligible time constant $\mu$. In this case,
which is also the one analyzed in section \ref{sec:2:sub:3} and
studied experimentally in section \ref{sec:4}, the control channel
will transform to
\begin{equation}\label{eq:actuator}
u(t) = v(t) - \mu \dot{u}(t),
\end{equation}
where $v(t)$ is the output value of a feedback controller is use.
Note that if the actuator time constant can be neglected, i.e.
$\mu = 0$, the system plant \eqref{eq:system}, \eqref{eq:actuator}
with the relative degree $r=3$ recovers to the one
\eqref{eq:system} with $r=2$.

The targeted control problem is to analyze the residual
stabilization control error $y(t)$ is response to unbiased (i.e.
zero mean-value) periodic disturbances with the known amplitude
$E$ and the upper bounded frequency range $\omega <
\omega_{\max}$. A nonlinear feedback control which uses the output
error, its time derivative, and integration is considered, while
the standard PID linear feedback control is to be compared with
it.

\subsection{CHOSM with proportional, derivative and integral terms}
\label{sec:2:sub:2}

The CHOSM control, as a simplified version of the so-called
Discontinuous Integral Controller \cite{cruz2019,moreno2020}, is
given by
\begin{equation}\label{eq:CSMC}
v = -k_1 \lceil y \rfloor^{1/3} - k_2 \lceil \dot{y} \rfloor^{1/2}
- k_3 \int \mathrm{sign}(y) dt,
\end{equation}
cf. \cite{perez2021,zamora2013control}. Here $k_1, k_2, k_3
> 0$ are the properly designed control gains. It can be
recognized that the corresponding three control terms are
equivalent to the proportional, derivative, and integral feedback
actions, respectively. The notation $\lceil x \rfloor^{p} \equiv
|x|^{p} \mathrm{sign}(x)$, for a system variable $x \in
\mathbb{R}$ and a number $0 \leq p \in \mathbb{R}$, is a commonly
used one, especially in the control related literature, so for
example $\lceil x \rfloor^{0} = \mathrm{sign}(x)$. The sign
function is defined as
\begin{equation}\label{eq:sign}
\mathrm{sign}(x) = \left\{%
\begin{array}{ll}
    1, & \hbox{ if } x>0;\\
    \left[-1,1\right], & \hbox{ if } x=0;\\
    -1, & \hbox{ if } x<0.
\end{array}%
\right.
\end{equation}
And the corresponding solutions $y(t)$ are in the Filippov sense
\cite{filippov1988}, meaning $y(t)$ is a locally absolutely
continuous function $y \, : \, [0, T) \rightarrow \mathbb{R}$ for
almost every $t \in (0,T)$.

\begin{figure}[!h]
\centering
\includegraphics[width=0.6\columnwidth]{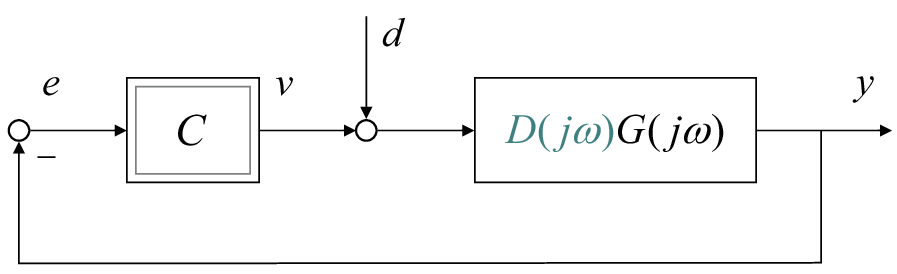}
\caption{Standard single-input-single-output (SISO) feedback
control loop with the objective of disturbance rejection.}
\label{fig1}
\end{figure}

For a perturbed second-order system, with the Lipschitz constant
$L > 0$ and the gains assignment $k_1 = \lambda_1 L^{2/3}$, $k_2 =
\lambda_2 L^{1/2}$, and $k_3 = \lambda_3 L$ according to
\cite{perez2021}, ensures the finite-time convergence and
insensibility of \eqref{eq:system} with respect to the matched
disturbances $d(t)$, cf. with Fig. \ref{fig1}. This is, however,
under assumption of no additional actuator dynamics $D(j\omega)$.
Further recall that the scaling factors $\lambda_1, \lambda_2,
\lambda_3 > 0$ can be assigned so as to ensure the finite-time
stability of \eqref{eq:system} with the known $L$, cf.
\cite{perez2021} and section \ref{sec:2:sub:3} below.

Note that in Fig. \ref{fig1}, a generic notation of the feedback
control $C$ refers to a transfer function $v(j\omega)/e(j\omega)$
in case of the linear PID-control, and to a continuous mapping $e
\in \mathbb{R} \mapsto v \in \mathbb{R}$ in case of the nonlinear
CHOSM control. The linear part of the system plant
\eqref{eq:system} is denoted by $G(j\omega)$, respectively.

\subsection{Harmonic balance based analysis}
\label{sec:2:sub:3}

One of the remarkable features of the nonlinear CHOSM control
\eqref{eq:CSMC}, that can be used for analyzing the loop
propagation of harmonic signals, is the closed analytic form of
the describing function (DF), as provided in \cite{perez2019b}
\begin{equation}\label{eq:DF}
N(A,\omega) = \frac{2\alpha_1 k_1}{\pi A^{2/3}} + j \,
\frac{2\alpha_2 k_2 \omega^{1/2}}{\pi A^{1/2}} + \Bigl( \frac{4
k_3}{\pi A} \Bigr) \Bigl( \frac{1}{j \omega} \Bigr).
\end{equation}
Here $\alpha_1 \approx 1.821$ and $\alpha_2 \approx 1.748$ are the
coefficients of the first harmonic approximation \cite{perez2021}.
Obviously, $j$ denotes the imaginary unit of the complex argument.
Since the DF method is based upon the filtering hypothesis, cf.
\cite{Atherton75}, it is imperative to assume that the CHOSM
control will operate on a plant which has low-pass
characteristics. This is fulfilled for the linear process
$D(j\omega)G(j\omega)$, cf. Fig. \ref{fig1}, where
$$
D(s) = (\mu s + 1)^{-1}.
$$
Then, the propagated harmonic in $y(t)$, and so in $e(t)$, allows
to compare the closed-loop characteristics including \eqref{eq:DF}
with those of a linear feedback controller $C(s)$, both at
steady-state.

Since the CHOSM control \eqref{eq:CSMC} provides a finite-time
convergence for the $d$-perturbed double integrator plant, cf.
Fig. \ref{fig1}, an additional actuator behavior in the loop, cf.
\eqref{eq:actuator}, can deteriorate convergence and induce
residual steady-state oscillations, also known as chattering. The
transfer properties of such servo system, i.e. from $d$ to $y$,
can be analyzed provided the external signal $d(t)$ is slow in
comparison with the self-excited oscillations. Then, the input can
be assumed approximately constant on one period of the fast
oscillations, cf. \cite{boiko2005}.

Here it is worth recalling that the fast self-excited oscillations
of the CHOSM controlled output can be preestimated as for the
corresponding amplitude and frequency, \cite{perez2021}:
\begin{equation}
\label{eq:AmplCSMC} \tilde{A} = L \Bigl( \frac{2\alpha_1
\lambda_1}{\pi \mathbb{K}^2 (1-\mathbb{K}^2)} \Bigr)^{3/2} \mu^{3}
\quad \hbox{ and } \quad \Omega = \frac{\mathbb{K}}{\mu}.
\end{equation}
That means the CHOSM controlled closed-loop system with $d=0$ and
actuator dynamics parameterized by $\mu$, cf. \eqref{eq:actuator},
will experience (after convergence) the residual steady-state
oscillations with frequency $\Omega$ and amplitude $\tilde{A}$.
The parameter $\mathbb{K} \in (0,1)$ is the solution of the
nonlinear equation, cf. \cite{perez2021},
\begin{equation}
\label{eq:KinChattering} 2 \mathbb{K} \lambda_1^{3/2} -
(1-\mathbb{K}^2)^{3/4} \lambda_2 \lambda_1^{3/4} + (1-
\mathbb{K}^2)^{3/2} \lambda_3 = 0.
\end{equation}
The closed-loop system (Fig. \ref{fig1}) with the CHOSM control
can be analyzed as two separate dynamic subsystems interacting
with each other via a set of parameters: the results of the
solution of the ''fast'' subsystem are used by the ''slow''
subsystem \cite{boiko2005}. This decomposition is possible if the
external input, i.e. $d(t)$, is much slower than the self-excited
oscillations, that is normally the case. That means for the
preestimated self-oscillations \eqref{eq:AmplCSMC} one needs to
guarantee $\omega_{\max} \ll \Omega$.

For analyzing propagation of an external disturbance
\begin{equation}
\label{eq:disturbance} d(t) = E \sin (\omega t + \phi_e)
\end{equation}
through the closed control loop (Fig. \ref{fig1}), we are to solve
the corresponding harmonic balance equation
\begin{equation}
\label{eq:hbe1}
\frac{1}{W(j\omega)}+N(A,\omega)=\frac{E}{A}e^{j(\phi_e - \phi)},
\end{equation}
which applies under the above separation assumption of the fast
(i.e. chattering) and slow (i.e. disturbance driven) system
dynamics. Here $W(j\omega)$ is the transfer function of the
overall linear system plant, including the actuator dynamics, and
$\phi$ is the output phase lag of steady-state oscillations at
$\omega$.

In the following, for the sake of simplicity and without loss of
generality, we assume a zero-phase unity disturbance, meaning
$E=1$ and $\phi_e = 0$. Note that for other $0 < E \neq 1$,
another $A(\omega)$-solutions of \eqref{eq:hbe1} need to be
determined each time, thus making the residual steady-state
oscillations in $y(t)$ both the frequency- and
amplitude-dependent, i.e. $A(\omega,E)$. We also assume an
actuator-perturbed linear plant given by
\begin{equation}
\label{eq:plantfrf}
W(j\omega)=\frac{K}{j\omega(j\omega\tau+1)(j\omega\mu+1)},
\end{equation}
where an additional first-order time delay element with the time
constant $\tau$ and input gain $K$ appears instead of the free
integrator, cf. \eqref{eq:system}, \eqref{eq:actuator}. Notice
that the plant transfer function \eqref{eq:plantfrf} coincides
with that of the experimental system investigated in this study,
cf. further with sections \ref{sec:3}, \ref{sec:4}. For obtaining
$A=A(\omega)$, we take first the absolute value
\begin{equation}
\label{eq:abs}
\left|\frac{1}{W(j\omega)}+N(A,\omega)\right|^2=\frac{1}{A^2}.
\end{equation}
Then, substituting \eqref{eq:plantfrf} and \eqref{eq:DF} into
\eqref{eq:abs}, one obtains the solution of $A$ in dependency of
$\omega$, and that for each given set of the numerical parameter
values.

\begin{figure}[!h]
\centering
\includegraphics[width=0.6\columnwidth]{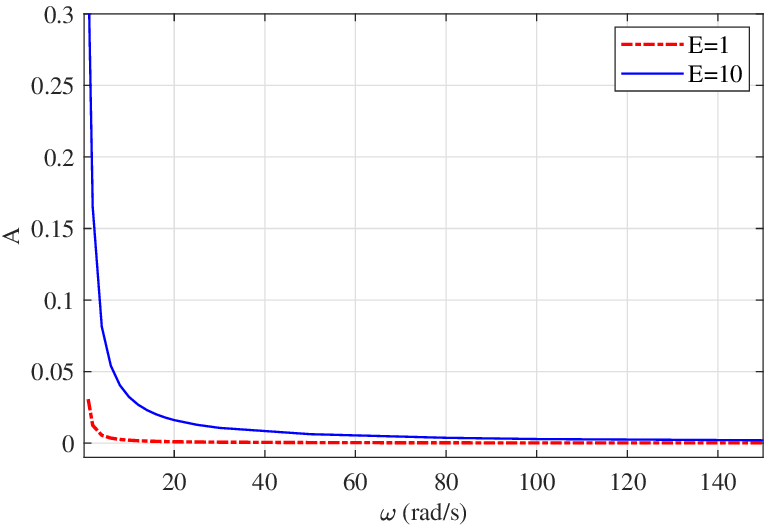}
\caption{Computed, based on \eqref{eq:absPart}, amplitude $A$ of
the output steady-state oscillations over the angular frequency
$\omega$ of the periodic disturbance, with $E=1$ (dash-dotted red
line) and $E=10$ (solid blue line).} \label{fig:AfromW}
\end{figure}

For the assigned $L=0.4$, the determined system parameters
$K=0.0408$, $\tau=0.0067$, $\mu=0.0012$ (cf. section \ref{sec:4}),
and the CHOSM control parameters $\lambda_1=2.7$,
$\lambda_2=5.345$, $\lambda_3=1.1$, which minimize the
oscillations amplitude \cite{perez2021}, the $A=A(\omega)$
solution is exemplary obtained by solving the equation
\begin{eqnarray}
\label{eq:absPart}
  A^{-2} &=& \left(\frac{1.69}{A^{2/3}} - 0.19\,\omega^2\right)^2 \\
\nonumber   & + & \left(24.5\,\omega \bigl(1 - 8.02\cdot
10^{-6}\omega^2 \bigr) +
  \frac{3.76\sqrt{\omega}}{\sqrt{A}} -
  \frac{0.56}{A\,\omega}\right)^2,
\end{eqnarray}
which results from \eqref{eq:abs} with \eqref{eq:plantfrf} and
\eqref{eq:DF}. Note that the otherwise unknown upper bound $L$ was
tuned based on the numerical simulations and experiments, so that
the residual oscillations of the output $y(t)$ at steady-state are
acceptably low. Numerical solution of \eqref{eq:absPart} obtained
by means of the Maple software yields the results visualized for
in Fig. \ref{fig:AfromW}, while the numerical solutions for $E=10$
are equally depicted for the sake of comparison.

%%%%%%%%%%%%%%%%%%%%%%%%%%%%%%%%%%%%%%%%%%%%%%%%%%%%%%%%%%%%%%%%%%%%%%%%%%%%%%%%
\section{Robust Linear PID Control}
\label{sec:3}

In this section, we provide a two-step procedure for robust PID
control design, which is resting on an underlying PD control with
the stable pole-zero cancelation and upper bound of the
disturbance sensitivity function.

In order to reduce the design complexity of a linear PID
controller and, correspondingly, to limit the degrees-of-freedoms
during the parameters tuning, a stable pole-zero cancelation is
first performed. Then, the tuning parameters reduce to only one
overall control gain and one integration time constant. To this
end, applying the underlying PD controller
\begin{equation}\label{eq:Cpd}
C_\mathrm{PD}(s) = \gamma(\tau s + 1),
\end{equation}
with the proportional gain $\gamma > 0$ and derivative time
constant $\tau > 0$, the dominant pole of the plant $\bar{G}(s)$
at $p = -1/\tau$ is canceled. Note that the performed linear
control design assumes the reduced plant model
\begin{equation}
\label{eq:redplantfrf} \bar{G}(s)=\frac{K}{s(\tau s +1)},
\end{equation}
cf. \eqref{eq:plantfrf}, while taking into account $\mu \ll \tau$.
Worth noting is that any small $\mu > 0$ will still create an
additional yet faster pole of the linear system plant and, thus,
inherently affect the resulting closed-loop dynamics. However,
this is minor for the proposed design which is based on the
sensitivity function, as becomes obvious from the developments
shown below. Following to that, only the $\gamma$-gain is used for
further PD control synthesis.

Since the disturbance rejection is of our major interest, the
input disturbance sensitivity function
\begin{equation}\label{eq:Syd}
S_\mathrm{yd}(s) = \frac{\bar{G}(s)}{1+\bar{G}(s)C(s)}
\end{equation}
is considered. Applying the PD controller \eqref{eq:Cpd} results
in
\begin{equation}\label{eq:SydPD}
S_\mathrm{yd,PD}(j\omega) = \frac{-K}{(\omega\tau-j)(\omega - j K
\gamma)},
\end{equation}
with the corresponding magnitude response
\begin{equation}\label{eq:SydPD_mag}
\left|S_\mathrm{yd,PD}(j\omega) \right| =
\frac{K}{\sqrt{\omega^2\tau^2+1}\sqrt{\omega^2 + K^2 \gamma^2}}.
\end{equation}
Note that it has its maximum at steady-state, i.e.
\begin{equation}
\Bigl|S_\mathrm{yd,PD}(j\omega)\Bigr|_{\omega = 0}  =
\frac{1}{\gamma}> \left|S_\mathrm{yd,PD}(j\omega) \right| \quad
\forall \: \omega>0,
\end{equation}
which is equivalent to
\begin{equation}\label{eq:Spd_hinf}
\left\|S_\mathrm{yd,PD}\right\|_\infty =  \frac{1}{\gamma} .
\end{equation}
Therefore, for a given worst-case amplification of the matched
disturbance, the proportional gain $\gamma$ is chosen as its
inverse.

Next, in order to ensure the steady-state accuracy, we make use of
a PID controller
\begin{equation}\label{eq:Cpid}
C_\mathrm{PID}(s) = \frac{\gamma (\tau s +1) (T_\mathrm{I}s +
1)}{T_\mathrm{I}s},
\end{equation}
with the integration time constant $T_\mathrm{I}$ to be the second
tuning parameter. This leads to the corresponding sensitivity
function~$S_\mathrm{yd,PID}$, cf. \eqref{eq:Syd}, with the
magnitude characteristics
\begin{equation}\label{eq:SydPID_mag}
\left|S_\mathrm{yd,PID}(j\omega)\right| = \frac{K
T_\mathrm{I}\omega}{\sqrt{(K \gamma - T_\mathrm{I} \omega)^2 + (K
T_\mathrm{I}\gamma\omega)^2} \sqrt{\tau^2\omega^2 + 1}}.
\end{equation}
In order to utilize the previous considerations (i.e. of
$S_\mathrm{yd,PD}$), we show that \eqref{eq:SydPID_mag} is bounded
by $\left\|S_\mathrm{yd,PD}\right\|_\infty$, i.e. that
\begin{equation}\label{eq:SydPID_gamma}
\left|S_\mathrm{yd,PID}(j\omega)\right| \leq \frac{1}{\gamma}\quad
\forall \: \omega>0
\end{equation}
holds true. This can be seen by considering
\begin{align*}
\left|S_\mathrm{yd,PID}(j\omega) \right|^2 &= \frac{(K T_\mathrm{I}\omega)^2}{\Bigl[ (K \gamma - T_\mathrm{I} \omega)^2 + (K T_\mathrm{I}\gamma\omega)^2\Bigr] \underbrace{(\tau^2\omega^2 + 1)}_{>1\,\forall\omega,\tau>0}}\\
&\leq \frac{(K T_\mathrm{I}\omega)^2}{(K \gamma - T_\mathrm{I}
\omega)^2 + (K T_\mathrm{I}\gamma\omega)^2} \equiv \Theta(j\omega)
\end{align*}
and knowing that $K,T_\mathrm{I},\omega >0$, which yields
\begin{equation*}
\Theta(j\omega) = \frac{1}{\left(\frac{K \gamma - T_\mathrm{I}
\omega}{K T_\mathrm{I}\omega} \right)^2 + \gamma^2} \leq
\frac{1}{\gamma^2}.
\end{equation*}
Therefore, the relation \eqref{eq:SydPID_gamma} becomes evident
and can be used for determining an appropriate proportional gain,
that for a specified minimal attenuation of the input disturbance.
The conservativeness of such design is discussed below in section
\ref{sec:4:sub:4}, that for the resulted system plant and $\gamma$
value.

The next free tuning parameter $T_\mathrm{I}$ is chosen based on
the frequency characteristics of the open-loop and the predefined
phase margin $\Phi$. The open-loop transfer function (with zero
initial conditions) is given by
\begin{equation}\label{eq:L}
H(s) = C_\mathrm{PID}(s) \, \bar{G}(s) = \frac{K\gamma
(T_\mathrm{I}s + 1)}{T_\mathrm{I}s^2},
\end{equation}
and consists of a double-integrator and one stable zero.
Therefore, the corresponding phase never crosses $-180^\circ$,
leading to a (theoretically) infinite gain margin. However, the
phase margin~$\Phi$ can be used as a further robustness measure,
especially with respect to an additional phase lag. Here we recall
that the plant transfer function $\bar{G}(s)$ does not take into
account an additional higher-frequent actuator dynamics.

Consider the amplitude and phase at the open-loop crossover
frequency $\omega_\mathrm{s}$ given by
\begin{subequations}\label{eq:L_omegas}
    \begin{align}
        \bigl| H(j\omega_\mathrm{s})\bigr| &=1\\
        \pi + \arg\bigl(H(j\omega_\mathrm{s})\bigr) &= \Phi .
    \end{align}
\end{subequations}
From \eqref{eq:L}, it can be seen that
\begin{equation*}
H(j\omega) = \frac{-K\gamma(1+j
T_\mathrm{I}\omega)}{T_\mathrm{I}\omega^2} =
-\frac{K\gamma}{T_\mathrm{I}\omega^2} - j \frac{K\gamma}{\omega}.
\end{equation*}
Hence, at $\omega_\mathrm{s}$ one obtains
\begin{subequations}\label{eq:L_omegas2}
    \begin{align}
    \bigl|H(j\omega_\mathrm{s})\bigr| &= \frac{K\gamma\sqrt{1 + \omega_\mathrm{s}^2 T_\mathrm{I}^2}}{\omega_\mathrm{s}^2 T_\mathrm{I}}\qquad\text{and}\\
    \arg\bigl(H(j\omega_\mathrm{s})\bigr) &= \arctan\left(T_\mathrm{I}\omega_\mathrm{s} \right) - \pi ,\qquad T_\mathrm{I},\omega_\mathrm{s} >0.
    \end{align}
\end{subequations}
The combination of \eqref{eq:L_omegas} and \eqref{eq:L_omegas2}
yields
\begin{subequations}\label{eq:L_omegas_results}
    \begin{align}
    \omega_\mathrm{s} &= \frac{K\gamma\sqrt{1 + \tan^2(\Phi)}}{\tan(\Phi)} \qquad\text{and}\\
    T_\mathrm{I} &=  \frac{\tan(\Phi)}{\omega_\mathrm{s}}.
    \end{align}
\end{subequations}

The above shown developments result in a two-step procedure for
the robust PID control design:
\begin{enumerate}
    \item Set a worst-case disturbance amplification $S_\mathrm{max}$ which is then leading to
        \begin{equation}
             \gamma=\frac{1}{S_\mathrm{max}}.
        \end{equation}
    \item Set the desired phase margin $\Phi$ and calculate
        \begin{equation}
            T_\mathrm{I} = \frac{\tan^2(\Phi)}{K\gamma\sqrt{1 + \tan^2(\Phi)}} .
        \end{equation}
\end{enumerate}
With the system and design parameters $\tau$, $\gamma$,
$T_\mathrm{I}$, the PID controller \eqref{eq:Cpid} is then
entirely determined.

%%%%%%%%%%%%%%%%%%%%%%%%%%%%%%%%%%%%%%%%%%%%%%%%%%%%%%%%%%%%%%%%%%%%%%%%%%%%%%%%
\section{Experimental Case Study}
\label{sec:4}

\subsection{System Setup}
\label{sec:4:sub:1}

\begin{figure}[!h]
\centering
\includegraphics[width=0.3\columnwidth]{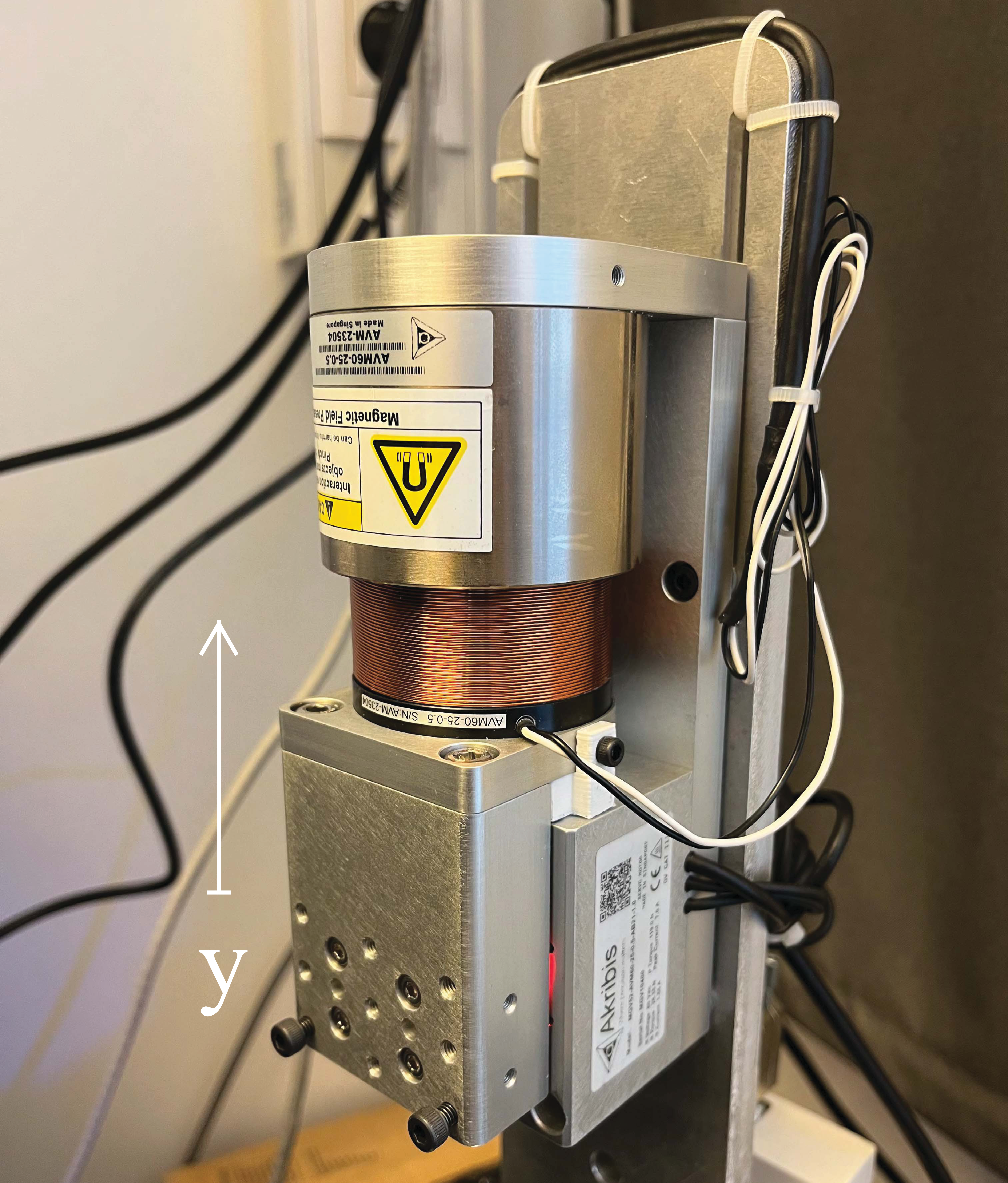}
\caption{Experimental setup of the motion system.} \label{fig:2:1}
\end{figure}

The second-order system under investigation is an
electro-mechanical linear-displacement actuator depicted in Fig.
\ref{fig:2:1}. The induced relative motion is indirectly measured
by the contactless inductive displacement sensor with a nominal
repeatability of $\pm 12$ micrometers. Despite all mechanical
elements are rigid, and the system dynamics is inherently of the
second-order, while having one free integrator, the nominal
electrical time constant $\mu = 1.2$ msec of the voice-coil-motor
is not fully negligible.
Furthermore, it is justified that the system has a relatively high
level of the sensor and process noise. The former is due to a
contactless sensing, while the latter is due to additional
parasitic by-effects which are not captured by the simplified
linear modeling, cf. sections \ref{sec:3} and \ref{sec:4:sub:2}.
The real-time control board operates the system with the set
sampling rate of 5 kHz. Further details on the experimental system
can be found in e.g. \cite{ruderman2022motion}. The partially
available and partially identified system parameters are given
below.

\subsection{Modeled and Identified System Plant}
\label{sec:4:sub:2}

The second-order dynamics of the system plant under consideration
is assumed to be represented by
\begin{equation}
\label{eq:systemplant}
    m\ddot{y}(t) + \sigma \dot{y}(t) = \frac{\Psi}{R} \, v(t) -
    mg,
\end{equation}
with input $v(t)$ (in volts, V) and output $y(t)$ (in meters, m).
The values of the moving mass $m$, electromotive force constant
$\Psi$, and coil resistance $R$ are given by
\begin{equation}
m = 0.538 \: \mathrm{kg},\quad \Psi =
17.16\,\mathrm{\frac{Vs}{m}},\quad R = 5.32\,\mathrm{\frac{V}{A}}
\end{equation}
and originate from the weight measurement and the actuator data
sheet \cite{AkribisVCM}. Note that the gravitational force
$F_\mathrm{g} = -mg$ with the gravity constant $g=9.81\,\mathrm{m
s^{-2}}$ can be directly compensated with $v_\mathrm{g} = mg R
\Psi^{-1}$ and is, therefore, not a part of further discussions.
In order to estimate the unknown damping parameter $\sigma>0$, we
apply $v + v_\mathrm{g}$ leading to
\begin{equation}
\frac{m}{\sigma}\ddot{y}(t) + \dot{y}(t) = \frac{\Psi}{R\sigma}
v(t).
\end{equation}
Using the Laplace transform, with zero initial conditions, the
transfer function of the system plant reads
\begin{equation}\label{eq:G}
\bar{G}(s) = \frac{K}{s(\tau s + 1)} \quad \hbox{with}  \quad
K=\frac{\Psi}{R\sigma}, \quad \tau=\frac{m}{\sigma}.
\end{equation}

Due to a free integrative behavior and bounded displacement $y\in
[0, \: 18]$ mm, the identification experiments are conducted in a
closed-loop, as in Fig. \ref{fig1}. In order to minimize the noise
amplification and not exciting further dynamics, a pure
proportional controller with a relatively low feedback gain
$\gamma = 100$ is utilized here for the parameters identification.
Starting from the middle of the displacement range, at
steady-state $y_0 = 9\,\mathrm{mm}$, a sinusoidal input
disturbance
\begin{equation}\label{eq:d}
d(t) = E \sin(\omega_\mathrm{d} \, t)
\end{equation}
is applied to $v(t)$ for exciting the closed-loop system. The
amplitude $E \in [0.7, \: 1.5]$ V is chosen to ensure a sufficient
excitation without input- and state-saturation, in accord with the
frequency range $\omega_\mathrm{d}\in [1, \: 1000]$ rad/s. A
measurement of 20 periods at steady-state is utilized for the
frequency response estimation at each frequency.
\begin{figure}[!ht]
    \centering
    \includegraphics[width=0.6\columnwidth]{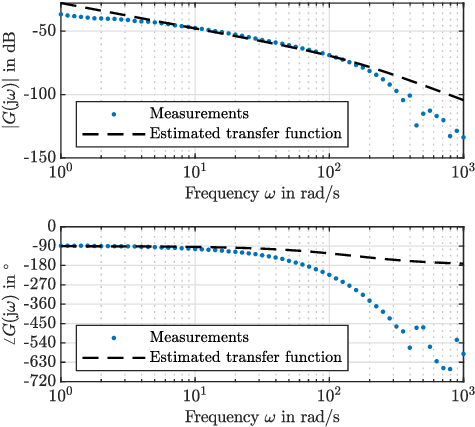}
    \caption{Bode plot of the experimental data and estimated transfer function.}
    \label{fig:2:2}
\end{figure}
With use of the auto-correlation $R_\mathrm{uu}(\cdot)$ and
cross-correlation functions $R_\mathrm{uy}(\cdot)$, the magnitude
and phase at $\omega_\mathrm{d}$ can be calculated as, cf. with
e.g. \cite{Isermann2011},
\begin{equation}
\bigl| \bar{G}(j\omega_\mathrm{d}) \bigr| =
\frac{R_\mathrm{uy,max}}{R_\mathrm{uu}(0)} \quad \text{and} \quad
\angle \bar{G}(j\omega_\mathrm{d}) = -\omega_\mathrm{d}\Delta\tau,
\end{equation}
where $\Delta \tau$ is the time lag between two correlation
functions. The resulting Bode plot of the experimentally collected
frequency-response data is depicted in Fig. \ref{fig:2:2}. The
phase at low frequencies corresponds to the integrating behavior.
However, high frequencies show an increasing time delay, in terms
of a phase lag, which is not covered by the second-order model
\eqref{eq:G}. Therefore, only the magnitude response is used for
the transfer function estimation. The least-squares optimal fit,
by using $\bigl| \bar{G}(j\omega_\mathrm{d}) \bigr|$ with
$\omega_\mathrm{d} \in [4, \: 380]$ rad/s, results in $\sigma =
80.49$. This yields the remaining parameter values
$$
K= 0.0408 \quad \text{and}\quad \tau = 6.684\;\mathrm{ms}.
$$
The Bode plot of the estimated $\bar{G}(j\omega)$ is also depicted
in Fig. \ref{fig:2:2} over the measurements.

\subsection{Robust Exact Differentiator}
\label{sec:4:sub:3}

In order to use the (otherwise) unavailable output derivative
$\dot{y}(t)$, we apply a robust exact differentiator
\cite{levant1998} (further as RED), which is based on the
sliding-mode estimation, see e.g. \cite{shtessel2014} for basics.
Its remarkable features are the insensitivity to a bounded noise,
provided the Lipschitz constant of n-th time-derivative is
available, and the finite-time convergence. This makes a RED
(theoretically) free of a phase lag. Recall that the latter is,
otherwise, deteriorating the closed-loop performance in case of a
low-pass filtering of the differentiated measurement of $y(t)$.
The second-order RED, cf. \cite{moreno2012}, with the
parametrization according to \cite{Reichhartinger2017}, is given
by
\begin{subequations}\label{eq:HOSM_diff}
\begin{align}
\dot{x}_0 &= x_1 + 3.1~ r~ \bigl| y-x_0 \bigr|^{\frac{2}{3}} ~\mathrm{sign}(y-x_0),\\
\dot{x}_1 &= x_2 + 3.2~ r^2~ \bigl| y-x_0 \bigr|^{\frac{1}{3}} ~\mathrm{sign}(y-x_0),\\
\dot{x}_2 &= 1.1 ~ r^3~ \mathrm{sign}(y-x_0),
\end{align}
\end{subequations}
where the scaling factor $r > 0$ is the single design parameter.
It is worth noting that $r^{n+1}$ (here $r^3$) corresponds to the
Lipschitz constant $L$ of the highest derivative ${y}^{(n)}$, cf.
\cite{Reichhartinger2017}. The second-order RED provides $x_0(t) =
y(t)$, $x_1(t) = \dot{y}(t)$, $x_2(t) = \ddot{y}(t)$, for all $t >
t_c$, where $t_c$ is the convergence time. Also we note that the
second-order (and not first-order) RED is purposefully used here,
in order to obtain a smoother estimate $x_1(t)$ of the output
derivative. The scaling factor $r=8$ was experimentally tuned on
the collected $y(t)$ data, for which an up-chirp signal until
$\omega = 80$ rad/s disclosed a still satisfying match between the
$x_1(t)$ estimate and the theoretically calculated $\dot{y}(t)$.

\begin{figure}[!ht]
    \centering
    \includegraphics[width=0.6\columnwidth]{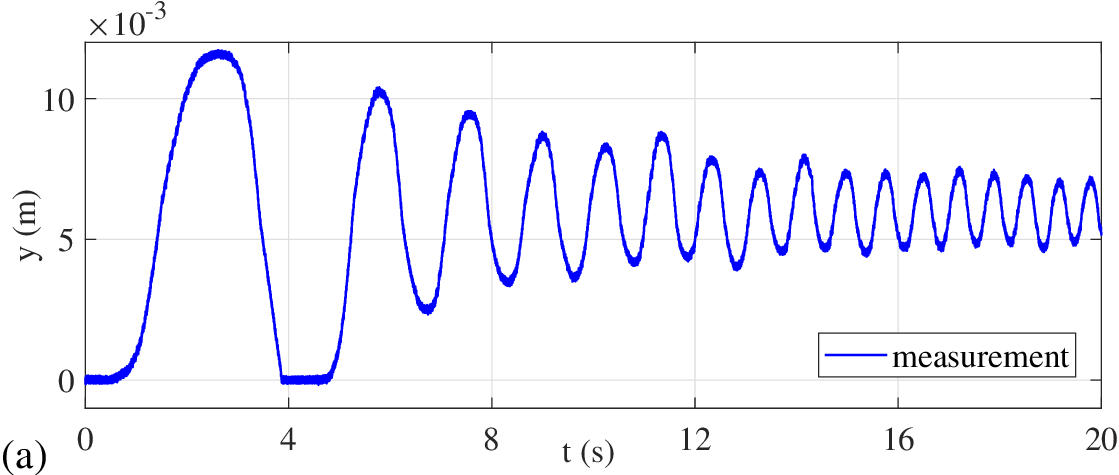}
    \includegraphics[width=0.6\columnwidth]{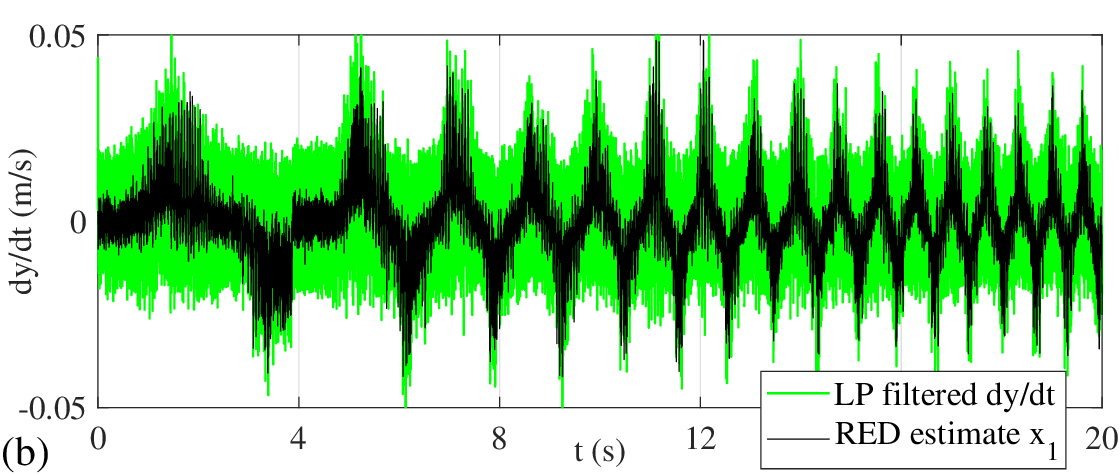}
    \caption{Exemplary evaluation of the 2nd order RED with $r=8$: the measured output value in (a) and
    RED estimate $x_1$ over the low-pass (LP) filtered discrete time derivative of $y$ in (b).}
    \label{fig:evalRED}
\end{figure}

An exemplary experimental evaluation of the RED
\eqref{eq:HOSM_diff} with $r=8$ is shown in Fig.
\ref{fig:evalRED}, in comparison with the discrete time derivative
of the measured signal $y(t)$ which is then low-pass filtered with
a cutoff frequency of 100 Hz.

\subsection{Synthesized PID controller}
\label{sec:4:sub:4}

\begin{figure}[ht]
    \centering
    \includegraphics[width=0.6\columnwidth]{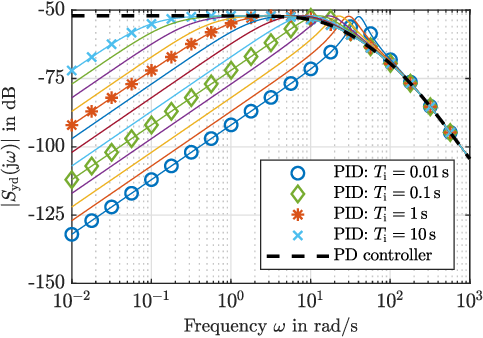}
    \caption{Magnitude plot of the input disturbance
    sensitivity function when using the PD controller opposite to PID
    controllers with the different integration time constants
    $T_\mathrm{I} \in [0.01, \: 10]$ rad/s.}
    \label{fig:Syd}
\end{figure}

\begin{figure}[ht]
    \centering
    \includegraphics[width=0.55\columnwidth]{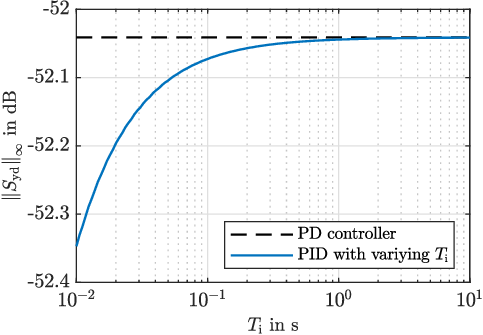}
    \caption{H-inf norms of the disturbance sensitivity functions from Fig. \ref{fig:Syd}.}
    \label{fig:Syd_maxima}
\end{figure}

The design approach developed in section \ref{sec:3} is used to
determine the PID controller for the identified plant $\bar{G}$.
For a sufficient disturbance rejection and robustness, we choose
\begin{equation*}
S_\mathrm{max}= -52\,\mathrm{dB}\quad \text{and} \quad\Phi =
60^\circ,
\end{equation*}
leading to
\begin{equation}\label{eq:PID_param_cascade}
\gamma = 400 \quad \text{and} \quad T_\mathrm{I} =
0.092\,\mathrm{s}.
\end{equation}
The magnitude plot of $S_\mathrm{yd}$, when using the underlying
PD controller, is depicted in Fig.~\ref{fig:Syd} opposite to the
use of PID controller, that for a large variation of the
integration time constant $T_\mathrm{I}$. Obviously, the maximum
disturbance amplification is only little dependent on the value of
$T_\mathrm{I}$, while a clear asymptotic reduction for lower
frequencies is evident. The PID controller with $T_\mathrm{I}=0.1$
s, cf. with the determined parameters
\eqref{eq:PID_param_cascade}, provides certain optimality in
shaping the disturbance sensitivity function. It yields without
sharp peak of $|S_\mathrm{yd}|$ and without a flat plateau of
$|S_\mathrm{yd}| \rightarrow S_{\max}$ upper bound which is
dictated by the underlying PD control.
A closer look at the H-inf norms for a variation of $T_\mathrm{I}$
in the relevant range, see Fig. \ref{fig:Syd_maxima}, confirms
\eqref{eq:Spd_hinf} and \eqref{eq:SydPID_gamma}. Worth noting here
is that for this set of parameters, the above mentioned
conservativeness of \eqref{eq:SydPID_gamma} is limited to
$0.31\,\mathrm{dB}$.

For ease of implementation and a better comparison with the
PID-like sliding-mode controller, the parallel form of PID
controller $C_\mathrm{PID}$ \eqref{eq:Cpid} reads as
\begin{equation}\label{eq:Cpid_parallel}
\bar{C}_\mathrm{PID}(s) = K_\mathrm{p} + K_\mathrm{I}\frac{1}{s} +
K_\mathrm{D}s.
\end{equation}
The proportional, integral, and derivative gains are
\begin{align}
K_\mathrm{p} &= \gamma\frac{T_\mathrm{I} + \tau}{T_\mathrm{I}} = 429.064,\\
K_\mathrm{I} &= \frac{\gamma}{T_\mathrm{I}} = 4348.267, \\
K_\mathrm{D} &= \gamma \tau = 2.674,
\end{align}
respectively. Note that the parameters result directly from
comparison of the coefficients between $C_\mathrm{PID}$ and
$\bar{C}_\mathrm{PID}$.

\begin{figure}[!ht]
    \centering
    \includegraphics[width=0.57\columnwidth]{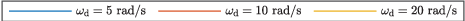}
    \includegraphics[width=0.6\columnwidth]{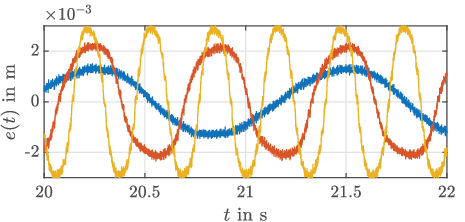}
    \caption{Measured control error $e$ of $\bar{C}_\mathrm{PID}$
    for the disturbance attenuation.}
    \label{fig:exp_HOSM}
\end{figure}

In order to evaluate efficiency of the proposed PID design, as
well as the accuracy of the model fit and the associated linear
loop shaping, cf. section \ref{sec:4:sub:2}, the sinusoidal input
disturbance $d(t)$, given by \eqref{eq:d}, is used in experiments.
The assigned amplitude is $E = 1\,\mathrm{V}$ and the frequencies
selected exemplary closer to $\omega_\mathrm{max}$ are
\begin{equation*}
\omega_\mathrm{d,1} = 5\,\mathrm{\frac{rad}{s}},\quad
\omega_\mathrm{d,2} =
10\,\mathrm{\frac{rad}{s}}\quad\text{and}\quad \omega_\mathrm{d,3}
= 20\,\mathrm{\frac{rad}{s}}.
\end{equation*}
The experimental results are depicted in Fig. \ref{fig:exp_HOSM}
by disclosing the control error $e = y - y_0$. Evaluating the
theoretically expected disturbance attenuation at
$\omega_\mathrm{d,1,\ldots,3}$ i.e.
$|\tilde{S}_\mathrm{yd}(j\omega_\mathrm{d,1,\ldots,3})| = \{-
58.6, \, -52.9, \, -53.8 \}$ dB and comparing these values with
the corresponding numbers calculated out from the experiments $\{-
57.5, \, -52.6, \, -50.1 \}$ dB, see Fig. \ref{fig:exp_HOSM}, one
can recognize a good accordance between both.

\subsection{Bode-like Loci of PID and CHOSM control}
\label{sec:4:sub:5}

\begin{figure}[!ht]
    \centering
    \includegraphics[width=0.6\columnwidth]{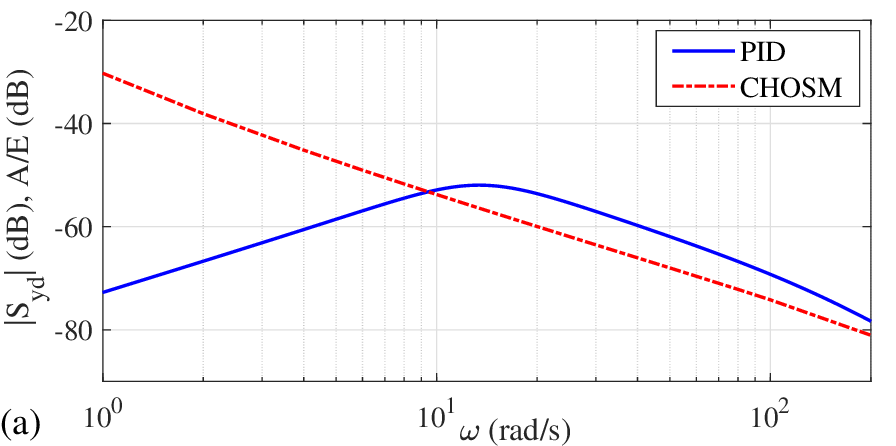}
    \includegraphics[width=0.6\columnwidth]{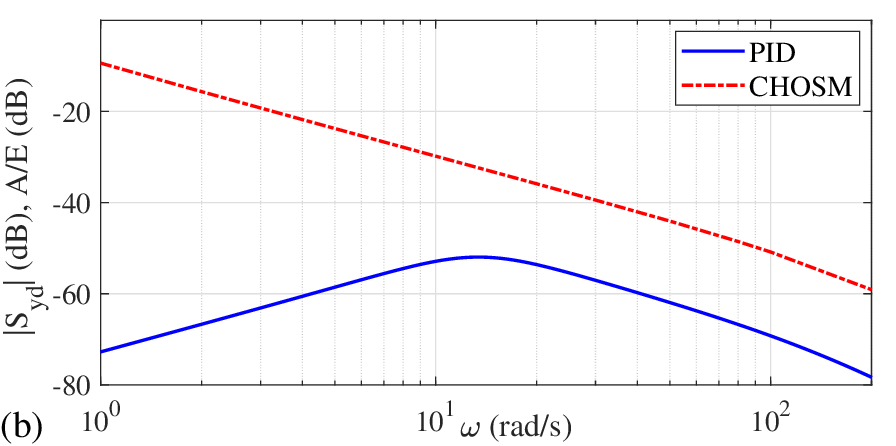}
    \caption{Amplitude response of $d$-to-$y$ frequency characteristics, for PID-controlled
    loop sensitivity function \eqref{eq:SydPID_mag}, and CHOSM-controlled harmonic balance
    \eqref{eq:abs}: for the disturbance amplitude $E=1$ in (a) and $E=10$ in (b).}
    \label{fig:sensbode}
\end{figure}

The frequency domain analysis of the disturbance sensitivity
characteristics at steady-state, developed in section \ref{sec:3}
for the linear PID-controlled and in section \ref{sec:2:sub:3} for
the nonlinear CHOSM-controlled closed-loop systems, yield both
comparable in terms of the Bode-like loci. Note that in case of
the CHOSM-controlled closed-loop system one can consider only the
Bode-like characteristics, correspondingly diagrams, since the
harmonic output response is not only frequency- but also
disturbance amplitude-dependent, i.e. $A(\omega, E)$. Moreover,
the available solution of \eqref{eq:abs}, allows regarding the
amplitude but not phase response, that is however sufficient for
analysis and comparison of the residual output errors. The
amplitude response of the $d$-to-$y$ characteristics is compared
in Fig. \ref{fig:sensbode} for the PID-controlled closed-loop
sensitivity function and CHOSM-controlled closed-loop harmonic
balance equation. The disturbance amplitude $E=1$, the same as
evaluated below in experiments, is assumed for the loci depicted
in (a), and amplitude $E=10$ for the loci depicted in (b).
One can recognize an advantage of the continuously decreasing $E$
to $A$ ratio of CHOSM with an increasing disturbance frequency,
cf. Fig. \ref{fig:AfromW}. On the contrary, the sensitivity
function of the PID-controlled closed-loop shows a typical
increase with a peak, first after which the disturbance to output
ratio starts to decrease as well.

\subsection{Controllers Comparison for Broadband Disturbances}
\label{sec:4:sub:6}

\begin{figure}[!ht]
    \centering
    \includegraphics[width=0.6\columnwidth]{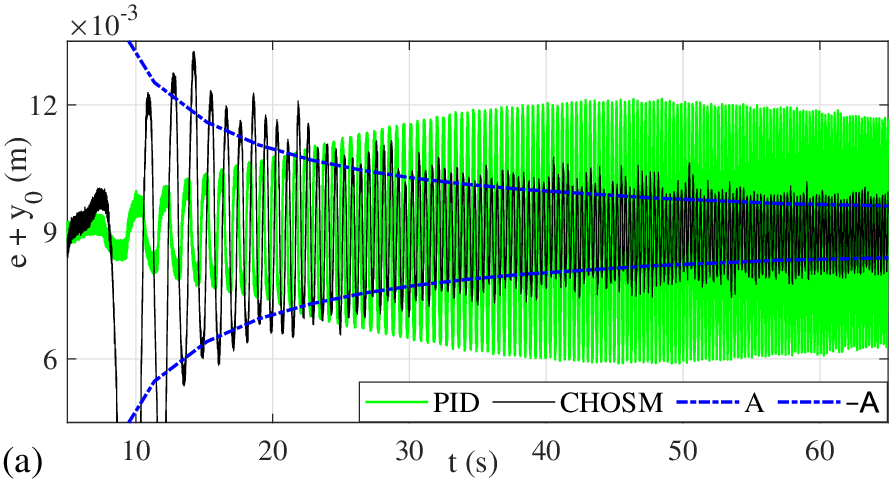}
    \includegraphics[width=0.6\columnwidth]{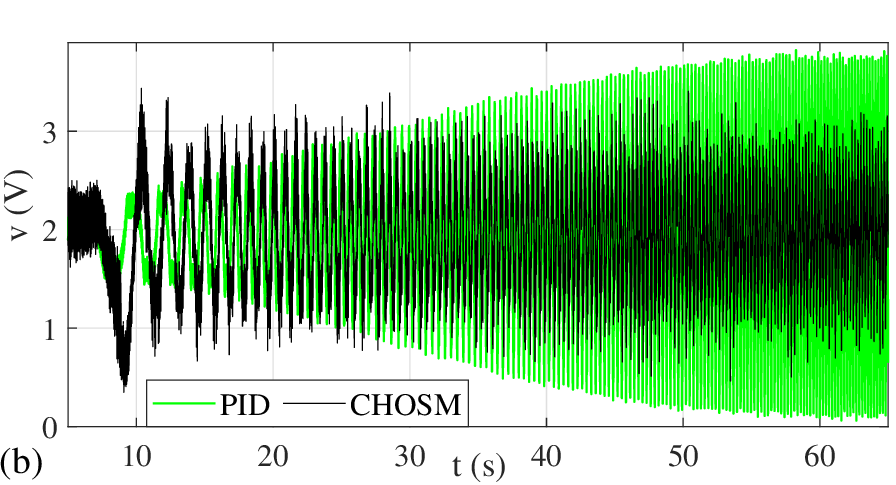}
    \caption{Experimental results of the PID and CHOSM controllers on rejection
    of the up-chirp disturbance with bandwidth 0.06--30 rad/s:
    the control error (around the reference position $y_0 = 9$ mm) in (a)
    and control signal in (b)}
    \label{fig:evalChrip}
\end{figure}

The design CHOSM and PID controllers, cf. sections \ref{sec:2} and
\ref{sec:3}, are evaluated experimentally and compared to each
other for the disturbance rejection. The applied disturbance value
constitutes an up-chirp signal
\begin{equation}\label{eq:chirpdist}
d(t) = E \sin \bigl((\omega_0 + \alpha t)\, t \bigr)
\end{equation}
with a linearly increasing frequency, i.e. $\alpha
> 0$, and the start and end frequencies at 0.06 rad/s and 30
rad/s, correspondingly. Note that the preestimated frequency of
the fast oscillations, see \eqref{eq:AmplCSMC},
\eqref{eq:KinChattering}, is $\Omega = 587$ rad/s, and thus the
requirement of $\omega_{\max} = 30 \ll \Omega$ is well fulfilled.
The chirp amplitude is $E=1$ V. The selected runtime of 65 s
ensures that the transient, correspondingly convergence, phase of
the control response is passed at already lower frequencies, cf.
Fig. \ref{fig:evalChrip} (a). Note that the convergence phase,
even though finite-time, is relatively long for the CHOSM control,
that for the range of feasible gain values. Following to that, a
steady-state disturbance rejection behavior can be compared for
the time frame about $t > 10$ s.
It can be seen that the error pattern of the PID control follows
the expected shape of the disturbance sensitivity function, cf.
Fig. \ref{fig:Syd}. On the contrary, the CHOSM control error is
continuously decreasing with an increasing frequency, that
corresponds to the CHOSM sensitivity function, cf. section
\ref{sec:4:sub:5}. The envelope $A(\omega)$, obtained by means of
solving the harmonic balance equation \eqref{eq:abs} and equally
plotted in Fig. \ref{fig:evalChrip} (a), fits very exactly with
the measured $y(t)$, which argues in favor of the developed
harmonic balance based analysis.

The period where the control error of both PID and CHOSM
controllers are approximately the same, i.e. at the time about 24
s, corresponds to the disturbance frequency $w_d \approx 9.5$
rad/s. Starting from that, the CHOSM control performs superior
comparing to the PID one in compensating for higher frequency
disturbances, that is well in accord with frequency
characteristics shown in Fig. \ref{fig:sensbode} (a).

Regarding the control values, see Fig. \ref{fig:evalChrip} (b),
one can recognize that for the PID control, the signal pattern is
also inline with the frequency response of the closed-loop system
(cf. Fig. \ref{fig1}). Here, the transfer characteristics between
$d(j \omega)$ and $v(j \omega)$ are indicative and can be directly
computed based on the transfer functions $\bar{G}(s)$ and
$\bar{C}(s)$. One can recognize that with an increasing angular
frequency, the amplitude of the PID control value $v(t)$ is also
growing, thus resulting in a higher power and, correspondingly,
energy consumption. On the contrary, the control value of the
CHOSM compensator keeps an almost constant amplitude level, that
speaks for a lower power consumption. Also recall that for
finite-time convergence of CHOSM controller, the control signal
tends to opposite value of the perturbation, that is the case for
chirp disturbance of a constant amplitude, cf. Fig.
\ref{fig:evalChrip} (b).

\begin{figure}[!ht]
    \centering
    \includegraphics[width=0.6\columnwidth]{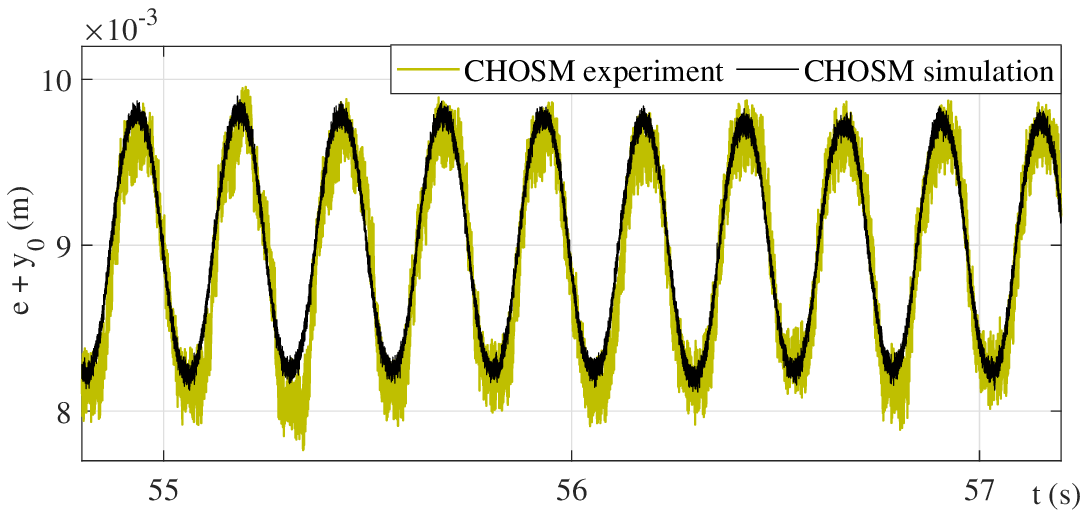}
    \caption{Comparison between the simulated and experimental response of
    CHOSM control; experimental data are from higher-frequencies from Fig. \ref{fig:evalChrip}.}
    \label{fig:evalCSMC}
\end{figure}

The comparison between the simulated and experimental response of
the closed loop with the CHOSM controller is further exemplary
shown in Fig. \ref{fig:evalCSMC}. Here the same experimental data,
a higher-frequency clipping from Fig. \ref{fig:evalChrip}, is
used. Note that in the numerical simulation of the system
\eqref{eq:systemplant}, augmented by the disturbance
\eqref{eq:chirpdist} and control \eqref{eq:CSMC}, an additional
band-limited white noise is added to the output $y(t)$, so as to
bring the behavior of the modeled control system closer to the
real one. Up to the noise level, the simulated and experimental
responses of the CHOSM-controlled output coincide well with each
other, see Fig. \ref{fig:evalCSMC}.

%%%%%%%%%%%%%%%%%%%%%%%%%%%%%%%%%%%%%%%%%%%%%%%%%%%%%%%%%%%%%%%%%%%%%%%%%%%%%%%%
\section{Conclusions}
\label{sec:5}

A PID-like continuous higher order sliding mode (CHOSM)
controller, known to be robust for rejection of the matched
Lipschitz continuous perturbations, was analyzed in terms of a
disturbance sensitivity function. Such frequency-domain approach,
which takes additionally into account the amplitude of
steady-state harmonics of the CHOSM control loop (also when
perturbed by an actuator) allows a comparison with standard linear
feedback controllers. Due to the structural similarity, i.e. with
feedback of the output, its derivative and integral, the CHOSM
control can and has been compared to a standard linear PID one,
designed also for rejection of the matched disturbances. Both
controllers were evaluated experimentally as the fair competitors.
The developed two-step procedure for a robust PID controller
design yields \emph{two tunable parameters}, the overall control
gain and the integration time constant. Both are shown associated
with the H-infinity norm of the disturbance sensitivity function
and with the desired phase margin, respectively. The CHOSM control
parametrization follows exactly the analysis and developments
provided in \cite{perez2021}, yielding only \emph{one tunable
parameter}, which is the scaling factor out from the Lipschitz
constant of perturbations. The comparative study of both
controllers is made for the second-order experimental motion
system with an additional (fast) parasitic dynamics of the
actuator. The two-parameters linear model of the system plant is
identified in frequency domain and shown to be sufficient for
designing the feedback controllers. For the time derivative of the
output, required for both PID and CHOSM control schemes, the
robust second-order sliding-mode differentiator is used in
experiments. The PID and CHOSM controllers are evaluated
experimentally on rejection of a broadband matched disturbance,
which is an up-chirp between 0.06 rad/s and 30 rad/s.

The resulted control performance is shown to be fully inline with
the theoretically expected (i) disturbance sensitivity
characteristics (i.e. sensitivity function) of the PID control
loop, and (ii) describing-function based prediction of the
steady-state harmonic oscillations of the CHOSM control loop. The
disturbance rejection performance of the CHOSM control is shown to
be clearly decreasing with an increasing frequency, as for the
control error pattern, and more energy efficient as for the
control signal amplitude. At the same time, the transient behavior
of the CHOSM control disclosed inferior comparing to the PID one,
that for the maximal achievable $L$-scaling factor of the CHOSM
control parametrization. The $L$-scaling factor, also related to
the actuator dynamics, proved to be most sensitive for application
of the CHOSM control. In summary, the demonstrated practical
comparison of PID and CHOSM controllers allows a better
understanding and distinction of the application benefits and
challenges of their use, when compensating for the amplitude- and
band-specific disturbances.

%%%%%%%%%%%%%%%%%%%%%%%%%%%%%%%%%%%%%%%%%%%%%%%%%%%%%%%%%%%%%%%%%%%%%%%%%%%%%%%%
\section*{Acknowledgments}
\label{sec:6} This work was financially supported by DAAD
scholarship programme: Research Stays for University Academics and
Scientists (DAAD ref. no. 91893483). Fourth author is
acknowledging support by CONACyT (Consejo Nacional de Ciencia y
Tecnologia) project 282013, PAPIIT-UNAM (Programa de Apoyo a
Proyectos de Investigacion e Innovacion Tecnologica) IN 115419.

\bibliographystyle{wileyNJD-AMS}
\bibliography{references}

\end{document}